\documentclass[pra,aps,twocolumn,showpacs]{revtex4}
\usepackage{graphicx}
\begin{document}
\title{Uhlmann's geometric phase in presence of isotropic 
decoherence}
\author{Jonas Tidstr\"{o}m$^{1,2}$\footnote{Electronic address: 
jonast@imit.kth.se} and 
Erik Sj\"{o}qvist$^{1}$\footnote{
Electronic address: erik.sjoqvist@kvac.uu.se}} 
\affiliation{$^{(1)}$Department of Quantum Chemistry, 
Uppsala University, Box 518, Se-751 20 Uppsala, Sweden \\ 
$^{(2)}$Department of Microelectronics and Information Technology, 
Se-164 40 Kista, Sweden}
\date{\today}
\begin{abstract}
Uhlmann's mixed state geometric phase [Rep. Math. Phys. {\bf 24}, 229
(1986)] is analyzed in the case of a qubit affected by isotropic
decoherence treated in the Markovian approximation. It is demonstrated
that this phase decreases rapidly with increasing decoherence rate and
that it is most fragile to weak decoherence for pure or
nearly pure initial states. In the unitary case, we compare Uhlmann's
geometric phase for mixed states with that occurring in standard
Mach-Zehnder interferometry [Phys. Rev. Lett.  {\bf 85}, 2845 (2000)]
and show that the latter is more robust to reduction in the length of
the Bloch vector. We also describe how Uhlmann's geometric phase in 
the present case could in principle be realized experimentally.
\end{abstract}
\maketitle

\section{Introduction}
The geometric phase, first discovered by Berry \cite{berry84} for
cyclic adiabatic quantal states and generalized to non-Abelian
\cite{wilczek84}, nonadiabatic \cite{aharonov87}, and noncyclic
\cite{samuel88} evolutions, has recently been suggested
\cite{zanardi99} and demonstrated experimentally
\cite{jones00} as a tool to achieve quantum computation that 
is resilient to certain types of errors. These analyses have 
been mainly concerned with the adiabatic geometric phase 
as this phase has the advantage that it depends only upon 
controllable external parameters \cite{xiang01}. However, 
from another point of view, adiabaticity is a serious 
drawback as it means that such gates operate slowly 
compared to the corresponding dynamical time scale 
and this makes them potentially vulnerable to loss of 
coherence. 

This issue has recently been addressed by Fonseca Romero {\it et al.}
\cite{fonseca02}, who have analyzed the behavior of the geometric phase 
in the presence of decoherence in the qubit (two-level) case and have
shown that the contribution from the adiabatic geometric phase to the
precession of the Bloch vector in a slowly rotating magnetic field is
robust to certain kinds of anistropic weak damping. Nazir {\it et al.}
\cite{nazir02} have analyzed implementations of geometric quantum
computation in the presence of decoherence and have demonstrated a
path-dependent sensitivity to anisotropic noise and loss of
entanglement for such gates.

A different perspective on the effect of decoherence arises 
when considering geometric phases for mixed states. Uhlmann 
\cite{uhlmann86} (see also Refs. 
\cite{uhlmann89,uhlmann91,hubner93,uhlmann96}) was probably
first to address this issue by lifting the density operator to a pure
state by attaching an ancilla and letting the combined system be
parallel transported along a specific purified path. More recently,
Sj\"{o}qvist {\it et al.} \cite{sjoqvist00} (see also Refs.
\cite{sjoqvist01,bhandari01,anandan01,slater01,sjoqvist02a,filipp02}) 
have discovered another mixed state geometric phase restricted 
to unitarily evolving density operators in the experimental 
context of one-particle interferometry. The geometric phases 
for mixed states proposed in \cite{uhlmann86} and \cite{sjoqvist00} 
are generically different in the unitary case and match only under 
very special conditions such as in the limit of pure states 
\cite{ericsson02a}.  

In this work, we analyze the behavior of Uhlmann's geometric phase in
the presence of decoherence and propose an experimental realization of
this phase. Specifically, we consider this geometric phase for a qubit
affected by isotropic depolarizing decoherence treated in the
Markovian approximation. The Markovian treatment of the depolarization
channel makes it possible to continuously monitor the mixed state
geometric phase for the Bloch vector in its motion towards the origin
inside the Bloch sphere. We hope that the present work would provide
insights into Uhlmann's mixed state geometric phase, extending the
purely mathematical treatments of this phase presented in the
literature so far to an explicit physical situation.

In the next section, we describe Uhlmann's approach to the mixed state
geometric phase. We apply the general formalism to the qubit case and
derive the corresponding parallel transport equation first obtained by
H\"{u}bner \cite{hubner93}. The unitary and Markovian approaches to
the depolarization channel are described in Sec. {\bf III}. Uhlmann's
geometric phase for a qubit evolution in the depolarization channel is
computed in Sec. {\bf IV}. In the idealized unitary pure state case,
this evolution describes an isosceles spherical triangle on the Bloch
sphere. We also describe how this geometric phase could in principle
be realized experimentally. The paper ends with the conclusions.

\section{Uhlmann's geometric phase}
The key idea in Uhlmann's \cite{uhlmann86} approach to the 
mixed state geometric phase is to lift the system's density 
operator acting on the Hilbert space $\mathcal{H}$ to an 
extended Hilbert space
\begin{equation}
{\cal H}^{ext} = {\cal H} \otimes {\cal H}^{a} 
\end{equation}
by attaching an ancilla $a$. This extension has the property 
that every unit vector (purification) $|\psi\rangle$ in the 
extended space may be reduced to a density operator in
the space of operators acting on ${\cal H}$ as  
\begin{equation}
\rho = \mbox{Tr} |\psi\rangle \langle \psi |,
\end{equation}
where $\mbox{Tr}$ is partial trace over the ancilla. 
The lift of $\rho$ is defined in terms of Hilbert-Schmidt 
operators $W:{\cal H}^{a} \longrightarrow {\cal H}$ such 
that $\rho = W W^\dag$. Such a lift is not unique as 
$W \mapsto \tilde{W} = WU$ generates the same $\rho$ for 
any choice of unitary $U$, i.e. we have a gauge freedom in 
the choice of $U$. Thus, we have an infinite number of ways 
to perform a purification of $\rho$. Our concern is now to 
pick out a class of exceptional purifications that defines 
a natural notion of geometric phase.

First, let us introduce the inner product between any pair 
$W_{1},W_{2}$ of Hilbert-Schmidt operators as  
\begin{equation}
\langle W_{1}, W_{2} \rangle \equiv 
\mbox{Tr} [W_{1}^{\dag} W_{2}] . 
\end{equation}
In terms of this we may define the quantity
\begin{eqnarray}
\nu = \langle \tilde{W}_{1}, \tilde{W}_{m} \rangle
\langle \tilde{W}_{m}, \tilde{W}_{m-1} \rangle \cdots 
\langle \tilde{W}_{2}, \tilde{W}_{1} \rangle 
\equiv \xi \, \langle \tilde{W}_{1}, \tilde{W}_{m} \rangle 
\nonumber \\ 
\end{eqnarray}
corresponding to the ordered set $\Pi=\rho_{1},...,\rho_{m}$ of 
density operators. Now, exceptional choices are the ones 
that maximize $|\xi|$, or, equivalently, that maximize each  
$|\langle \tilde{W}_{j+1}, \tilde{W}_{j} \rangle |$. 

By making use of $W=\sqrt{\rho} V$, $V$ being unitary, and 
the polar decomposition $\sqrt{\rho}_{j+1} \sqrt{\rho}_j = 
|\sqrt{\rho}_{j+1} \sqrt{\rho}_j| U_{j+1,j}$ we obtain a 
necessary and sufficient condition for $|\xi|$ maximal 
when 
\begin{equation}
|\langle \tilde{W}_{j+1}, \tilde{W}_j \rangle| = 
\mbox{Tr} \Big[ \sqrt{\sqrt{\rho}_{j+1} \rho_j 
\sqrt{\rho}_{j+1}}\, \Big] . 
\end{equation}
It follows that $ \nu=\xi \, \langle \tilde{W}_{1}, 
\tilde{W}_{m} \rangle$ is invariant under the remaining 
gauge freedom $\tilde{W}_j \mapsto \epsilon_jU\tilde{W}_j$, 
$|\epsilon_j|=1$ and  $U$ a fixed unitarity. Hence, with 
$\tilde{W}$ being exceptional, $\nu$ depends only upon the 
ordered set $\Pi$ and we define the corresponding Uhlmann phase 
as $\phi_{g} [\Pi] \equiv \arg \nu$. This phase is real-valued, 
gauge invariant, and, in the $m\longrightarrow \infty$ limit, 
independent of the subdivision of the path. These properties 
make $\phi_{g}[\Pi]$ a natural definition of the mixed state 
geometric phase associated with $\Pi$. Furthermore, by 
introducing $\phi_{j+1,j} = \arg \langle W_{j+1},W_{j} \rangle$ 
we may write    
\begin{eqnarray}
\phi_g[\Pi] & = 
\phi_{m, m-1}+\cdots+\phi_{3,2}+\phi_{2,1}+\arg\langle \tilde{W}_1,
\tilde{W}_m \rangle . 
\nonumber \\ 
\label{eq:gengauge} 
\end{eqnarray} 
This may be simplified by choosing a particular class of gauge 
that corresponds to parallel lift defined by requiring 
\begin{equation}
\tilde{W}_{j+1}^{\dag} \tilde{W}_j > 0, \ \forall j, 
\label{eq:parallellift} 
\end{equation} 
which implies $\phi_{j+1,j}=0$ for each intermediate step 
between $1$ and $m$ in Eq. (\ref{eq:gengauge}). Note that 
the main condition, i.e. that $|\xi|$ is maximal, is still 
true when Eq. (\ref{eq:parallellift}) holds. For this 
important class of gauge Uhlmann's mixed state geometric 
phase reads 
\begin{equation} 
\phi_g [\Pi] = \arg\langle \tilde{W}_1, \tilde{W}_m \rangle . 
\label{eq:puhlphase} 
\end{equation} 

Let us now consider the parallel transport condition in the particular
case of a qubit (two-level system). Any qubit state can be written as
$\rho = \frac{1}{2} (1+{\bf r} \cdot \mbox{\boldmath $\sigma$})$,
where ${\bf r}$ is the Bloch vector, the length of which being less
than unity for mixed states, and $\mbox{\boldmath $\sigma$} =
(\sigma_{x},\sigma_{y},\sigma_{z})$ are the standard Pauli 
matrices. The parallelity lift condition Eq. (\ref{eq:parallellift})
for any pair $\rho_{j}$ and $\rho_{j+1}$ of such mixed qubit states
\cite{remark1} is equivalent to
\begin{equation}
V_{j+1} V_{j}^{\dag}=\sqrt{\rho^{-1}_{j+1}}\sqrt{\rho^{-1}_{j}}
\sqrt{\sqrt{\rho_{j}}\rho_{j+1}\sqrt{\rho_{j}}} . 
\end{equation}  
By writing $\sqrt{\rho_{j}} = a_0 + {\bf a} \cdot 
\mbox{\boldmath $\sigma$}$ and $\sqrt{\rho_{j+1}} = b_0 + 
{\bf b} \cdot\mbox{\boldmath $\sigma$}$, and using 
features of $2\times2$ matrices we obtain \cite{hubner93} 
\begin{equation}
V_{j+1} V_{j}^{\dagger} = \frac{b_0 a_0 + {\bf b} \cdot {\bf a} +
i ({\bf b} \times {\bf a}) \cdot \mbox{\boldmath $\sigma$}}
{\sqrt{(b_0 a_0 + {\bf b} \cdot {\bf a})^2 + 
|{\bf b} \times {\bf a}|^2}}.
\label{eq:v2v1}
\end{equation}
For a continuous path $\Pi : t\in [0,\tau] \mapsto {\bf r} (t)$ 
of qubit states we need to consider the infinitesimal version 
of Eq. (\ref{eq:v2v1}). Let $V_{j}=V$, $V_{j+1}=V+dV$, 
$\sqrt{\rho_{j+1}} = \sqrt{\rho_{j}}+d\sqrt{\rho_{j}}$ 
and use that $a_{0}^{2} + |{\bf a}|^2 = \frac{1}{2}$ we 
obtain to first order in $d{\bf a}$ the differential equation 
\cite{hubner93}
\begin{equation}
dVV^{\dag} = 2i(d{\bf a} \times {\bf a}) \cdot 
\mbox{\boldmath $\sigma$} . 
\label{eq:etbs}
\end{equation} 
Formally we may write the solution of Eq. (\ref{eq:etbs}) as 
\begin{equation}
V = {\cal P} \exp \Big( 2i \int ( d{\bf a} \times {\bf a} ) 
\cdot \mbox{\boldmath $\sigma$} \Big) V_{0} , 
\label{eq:time}
\end{equation}
where ${\cal P}$ stands for path ordering and $V_{0}$ is the 
initial unitarity. Evaluating this path ordered expression and 
inserting into Eq. (\ref{eq:puhlphase}) yields Uhlmann's mixed 
state geometric phase for any qubit state. 

In the qubit case, it has been argued \cite{hubner92,braunstein95}
that the interior of the Bloch sphere is curved. This result is
essentially captured by the mixed state line element
\cite{braunstein95}
\begin{equation}
ds^2 = \frac{dr^2}{1-r^2} + r^2 d{\bf n}\cdot d{\bf n} 
\label{eq:linelement}
\end{equation} 
with ${\bf r}=r{\bf n}$, which shows that as one moves away from the
center at $r=0$, the circumference of the 2-sphere defined by each
fixed $r>0$ grows more slowly than the distance from the center, due
to the factor $1/(1-r^2)$ in front of $dr^{2}$. Uhlmann's geometric
phase for a qubit is the holonomy that naturally measures this
curvature inside the Bloch sphere, just as the standard pure state
geometric phase is the holonomy that measures the curvature of the
Bloch sphere. In view of this and due to the non-Euclidean behavior of
the radius and circumference when moving away from the origin inside
the Bloch sphere, there is no reason to expect that the Uhlmann phase
should have any direct relation to the solid angle enclosed by the
Bloch vector in three dimensional Euclidean space. The result of the
calculation in Sec. {\bf IV} below for a qubit in the depolarization
channel may be regarded as an illustration of this intuitive
reasoning.

\section{Depolarization channel}
In the depolarization channel, the environment induces 
isotropic errors in the qubit state. This may be 
represented by taking the three errors 
\begin{itemize}
\item[(i)]  $|\psi\rangle \mapsto \sigma_x |\psi\rangle$,  
bit flip, 
\item[(ii)] $|\psi\rangle \mapsto \sigma_z|\psi\rangle$, 
phase flip, 
\item[(iii)] $|\psi\rangle \mapsto \sigma_y|\psi\rangle$, 
both bit and phase flip, 
\end{itemize}
to be equally likely, each occurring with probability $p/3$.
A unitary representation of the channel, using a minimal set 
of ancilla states $\{ |\tilde{0}_{a} \rangle, \ldots , 
|\tilde{3}_{a} \rangle \}$, is given by 
\begin{eqnarray} 
U & : & \rho \otimes |\tilde{0}_{a} \rangle \langle \tilde{0}_{a}| 
\longrightarrow \left[ \sqrt{1-p} |\tilde{0}_{a} \rangle \right. 
\otimes I 
\nonumber \\ 
 & & + \sqrt{\frac{p}{3}} \Big( |\tilde{1}_{a} \rangle 
\otimes \sigma_{x} + |\tilde{2}_{a} \rangle \otimes \sigma_{y}
+ \left. |\tilde{3}_{a} \rangle \otimes \sigma_{z} \Big) 
\right] 
\nonumber \\ 
 & & \times R \rho R^{\dagger} 
\left[ \sqrt{1-p} I \otimes \right. 
\langle \tilde{0}_{a}| 
\nonumber \\ 
 & & + \sqrt{\frac{p}{3}} \Big( \sigma_{x} 
\otimes \langle \tilde{1}_{a}| + \sigma_{y} \otimes \langle 
\tilde{2}_{a}|
\left. + \sigma_{z} \otimes \langle \tilde{3}_{a}| 
\Big) \right] , 
\nonumber \\ 
\label{eq:unitaryrep}  
\end{eqnarray} 
where $R$ is some unitary operation acting only on the qubit.
This may be lifted into the pure state evolution by adding 
another ancilla system $b$ such that the initial state reads  
\begin{eqnarray}
\rho \otimes |\tilde{0}_{a} \rangle \langle \tilde{0}_{a}|
 & = & \left( \frac{1+r}{2} |\psi \rangle \langle \psi | \right. 
\nonumber \\ 
 & & \left. + \frac{1-r}{2} |\psi^{\perp} \rangle 
\langle \psi^{\perp} | \right) \otimes |\tilde{0}_{a} \rangle 
\langle \tilde{0}_{a}| 
\nonumber \\ 
 & \longrightarrow & |\Psi \rangle = \left( \sqrt{\frac{1+r}{2}} 
|\psi \rangle \otimes |0_{b} \rangle \right. 
\nonumber \\ 
 & & \left. + \sqrt{\frac{1-r}{2}} 
|\psi^{\perp} \rangle \otimes |1_{b} \rangle \right) \otimes 
|\tilde{0}_{a} \rangle 
\nonumber \\   
\label{eq:purdepol}  
\end{eqnarray}
and extending $U$ to $U\otimes I_{b}$. 

Similarly, under certain restrictions one may model the map of 
the system's density operator as a nonunitary Markovian evolution 
described by the Lindblad equation \cite{lindblad74} ($\hbar=1$ 
from now on)
\begin{eqnarray}
\dot{\rho} = 
i[\rho,H]+\sum_{\mu} \Big( L_{\mu} \rho L_{\mu}^{\dagger} -
\frac{1}{2} L_{\mu}^{\dagger} L_{\mu} \rho - \frac{1}{2} 
\rho L_{\mu}^{\dagger} L_{\mu} \Big) . 
\nonumber \\ 
\label{eq:buff}
\end{eqnarray}
Each term $L_\mu\rho L_\mu^{\dag}$ represents one of the possible 
errors (quantum jumps), while the sum over $-\frac{1}{2} L_\mu^{\dag} 
L_\mu\rho - \frac{1}{2}\rho L_\mu^{\dag}L_\mu$ is needed to satisfy 
the normalization condition. In this framework, the depolarization 
channel may be represented by the Lindblad operators  
\begin{equation}
L_{\mu}=\sqrt{\frac{\Gamma}{3}}\sigma_{\mu}, \quad \mu=1,2,3,
\end{equation} 
where $\Gamma$ is the time-independent decoherence rate. Inserting 
these $L_{\mu}$'s into Eq. (\ref{eq:buff}), we obtain the Lindblad 
equation for the depolarization channel as
\begin{equation}
\dot{\rho} = i[\rho,H] - \frac{2\Gamma}{3} {\bf r} 
\cdot\mbox{\boldmath $\sigma$} . 
\label{eq:depoleq}
\end{equation}
Assuming the Hamiltonian $H = \frac{1}{2} \omega \sigma_z$,  
corresponding to the unitarity $R = \exp \big( -\frac{i}{2} 
\omega t \sigma_{z} \big)$, and the initial condition 
${\bf r} (0) = r_{0} (\sin\theta,0,\cos\theta)$, the solution 
of Eq. (\ref{eq:depoleq}) reads 
\begin{equation}
{\bf r} (t) = r_{0} \ e^{-\frac{4\Gamma}{3}t}
(\sin \theta \cos \omega t , \sin \theta \sin \omega t , 
\cos \theta ) .
\label{eq:blocheq}
\end{equation}
Thus, the Bloch vector precesses uniformly around the $z$ 
axis and its length decreases isotropically. The effects 
of the unitary and Markovian treatments of the depolarizing 
channel can be formally related at time $t$ as $p(t) = 
\frac{3}{4} (1-r(t)/r_{0}) = \frac{3}{4} (1-e^{-4\Gamma t/3})$.  

\section{Uhlmann's geometric phase in the depolarization channel}
\subsection{Theoretical analysis}
To evaluate Eq. (\ref{eq:time}) analytically in the depolarization 
channel, we may choose a path where $( d{\bf a} \times {\bf a} ) 
\cdot \mbox{\boldmath $\sigma$}$ is time-dependent in such a way 
that it commutes at different times within a set of time intervals. 
Such a path is  
\begin{eqnarray}
A \rightarrow B & : & {\bf r} (t) = 
r_{0} \ e^{-\frac{4\Gamma}{3}t} (\sin\omega t, 0,\cos\omega t), 
\nonumber \\ 
 & & 0 \le t \le \pi /(2 \omega) ,  
\nonumber\\
B \rightarrow C & : & {\bf r} (t) = 
r_{0} \ e^{-\frac{4\Gamma}{3}t} (\sin\omega t, -\cos \omega t, 0), 
\nonumber \\ 
 & & \pi/(2 \omega) \le t \leq (\varphi + \pi/2)\omega, 
\nonumber\\
C \rightarrow D & : & {\bf r} (t) = 
r_{0} \ e^{-\frac{4\Gamma}{3}t} 
(\cos \varphi \sin [\omega t -\varphi ] , 
\nonumber \\ 
 & & \sin \varphi \sin [\omega t - \varphi ] , 
-\cos [\omega t - \varphi]), 
\nonumber \\  
 & & (\varphi + \pi /2) /\omega \le t \le (\varphi + \pi )/ \omega , 
\label{eq:path}
\end{eqnarray} 
which is shown in Fig. 1 in the particular case where 
$\varphi = \pi/2$. 

The idealized pure state evolution is obtained when $\Gamma =0$ and
$r_{0} = 1$; it is represented by the solid line in Fig. 1 that
defines the geometric phase $-\pi /4$. In the general case,
the third rotation is taken around the direction ${\bf m} = (\sin
\varphi, -\cos \varphi ,0)$ and the pure state geometric phase becomes
$-\varphi /2$.

\begin{figure}[ht!]
\begin{center}
\includegraphics[width=8 cm]{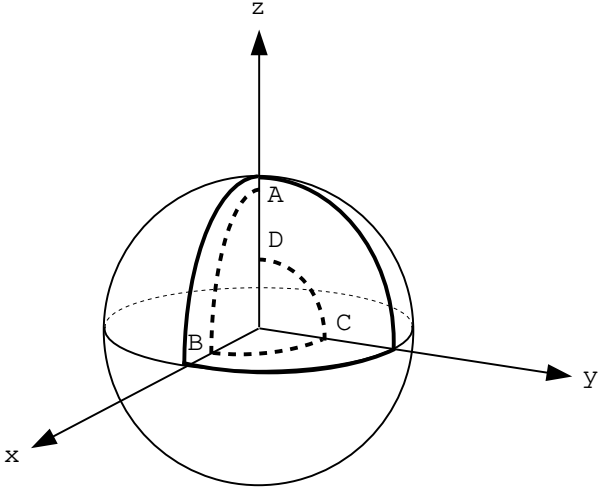}
\end{center}
\caption{Path with istropically decreasing length of the Bloch 
vector. The solid line represents the unitary pure state case 
and defines a spherical triangle enclosing the solid angle 
$\pi /2$. The dashed line represents the case with nonvanishing 
decoherence.}
\end{figure}

To compute $V_{D}$ and thereby Uhlmann's mixed state geometric 
phase for any fixed $\Gamma$, we first note that with $a(t) = 
|{\bf a} (t)|$ it follows that $d{\bf a} \times {\bf a} = 
a^2(t) d{\bf n} \times {\bf n}$ and for the above path we have 
\begin{eqnarray}
A \rightarrow B & : & d{\bf n} \times {\bf n} = 
-\omega dt \, {\bf y} , 
\nonumber\\
B \rightarrow C & : & d{\bf n} \times {\bf n} = 
-\omega dt \, {\bf z} , 
\nonumber\\
C \rightarrow D & : & d{\bf n} \times {\bf n} = 
-\omega dt \, {\bf m} ,
\end{eqnarray}
where ${\bf y}$, ${\bf z}$, and ${\bf m}$ are unit vectors 
along the three rotation axes. Substituting this into Eq. 
(\ref{eq:time}) yields 
\begin{eqnarray}
V_D & = & \exp \Big( -2i\omega 
\int_{(\varphi + \pi/2) /\omega}^{(\varphi + \pi)/ \omega)} 
a^{2}(t) \, dt \ {\bf m} \cdot \mbox{\boldmath $\sigma$} \Big) 
\nonumber \\ 
 & & \times \exp \Big( -2i\omega 
\int_{\pi /(2\omega)}^{(\varphi + \pi/2) /\omega} 
a^{2}(t) \, dt \ \sigma_{z}  \Big) 
\nonumber \\ 
 & & \times \exp \Big( -2i \omega \int_{0}^{\pi /(2\omega)} 
a^2 (t)\, dt \ \sigma_{y} \Big) V_A 
\nonumber \\ 
 & \equiv & e^{-i\chi{\bf m} \cdot \mbox{\boldmath $\sigma$}} 
e^{-i\kappa \sigma_{z}} e^{-i\mu \sigma_{y}}V_A , 
\label{eq:vd}
\end{eqnarray}
where we have used that $V_{0} = V_{A}$. Now, from 
$(\sqrt{\rho})^{2} = \rho$ and $\sqrt{\rho} = a_{0} + a 
{\bf n} \cdot \mbox{\boldmath $\sigma$}$, we obtain $a^2 (t) = 
\left( 1-\sqrt{1-r^2(t)} \, \right) /4$ so that the angles $\mu$, 
$\kappa$, and $\chi$ have the generic form 
\begin{eqnarray} 
 & 2\omega & \int_{t_{k}}^{t_{k}+\Delta \varphi /\omega}  
a^2(t) dt = \frac{\Delta \varphi}{2} 
\nonumber \\ 
 & & - \frac{\omega}{2} 
\int_{t_{k}}^{t_{k}+\Delta \varphi /\omega} 
\sqrt{1-r_{0}^{2} e^{-\frac{8\Gamma}{3}t}} \ dt  
\end{eqnarray}
with $t_{k} = 0,\pi/(2 \omega),[\varphi + \pi/2] /\omega$ and 
$\Delta \varphi = \pi/2 , \varphi , \pi/2$, respectively. 
The integral can be evaluated by using 
\begin{eqnarray}
 & & \frac{8\Gamma}{3} \int \sqrt{1 - r_{0}^{2} 
e^{-\frac{8\Gamma}{3} t}} dt = -2\sqrt{1 - r_{0}^{2} 
e^{-\frac{8\Gamma}{3} t}} 
\nonumber \\ 
 & & + \ln \left( \frac{1 + \sqrt{1 - 
r_{0}^{2} e^{-\frac{8\Gamma}{3} t}}}{1-\sqrt{1 - r_{0}^{2} 
e^{-\frac{8\Gamma}{3} t}}} \right) = 2 \Big[ 4a^{2} (t) - 1 + 
\ln \frac{a_{0} (t)}{a(t)} \Big] 
\nonumber \\ 
\label{eq:integral}
\end{eqnarray} 
with the concomitant integration limits for $\mu ,\kappa$,  
and $\chi$. Let us further introduce 
\begin{eqnarray}
\sqrt{\rho_A} & = & a_{0} (0) + a(0) \sigma_{z} \equiv 
\alpha + \beta \sigma_z ,  
\nonumber \\
\sqrt{\rho_D} & = & 
a_{0} \left( \frac{\varphi + \pi}{\omega} \right) + 
a\left( \frac{\varphi + \pi}{\omega} \right) \sigma_{z} 
\nonumber \\ 
 & \equiv & \nu + \eta \sigma_z . 
\label{eq:deffa}
\end{eqnarray}
In terms of Eqs. (\ref{eq:vd}) and (\ref{eq:deffa}) the 
geometric phase for the path $A \rightarrow B \rightarrow 
C \rightarrow D$ becomes 
\begin{eqnarray}
\phi_g & = & \arg \mbox{Tr} [\sqrt{\rho_{A}} \sqrt{\rho_{D}} 
V_{D} V_{A}^{\dagger} ] 
\nonumber \\ 
 & = & -\arctan  \left[ 
\frac{\alpha \eta + \beta \nu}{\alpha \nu + \beta \eta} \right. 
\nonumber \\ 
 & & \times \left. \left( \frac{\sin \kappa + 
\sin [\varphi - \kappa] \tan \chi \tan \mu}{\cos \kappa + 
\cos [\varphi - \kappa] \tan \chi \tan \mu} \right) \right] . 
\label{eq:fjek} 
\end{eqnarray} 

To further analyze this general analytic result, let us 
consider the following important special cases: 

\begin{enumerate} 

\item {\it Pure and unitary case:} Here, $r(t) =1, \ \forall 
\ t>0$, which implies that $\alpha = \beta = \nu = \eta = 
\frac{1}{2}$, $\mu = \chi = \pi /4$, and $\kappa = \varphi /2$ 
yielding $\phi_g = -\varphi /2$. This   is the expected pure 
state geometric phase for a qubit enclosing the solid angle 
$\varphi$ on the Bloch sphere.   

\item {\it Mixed and unitary case:} Here, $\Gamma = 0$ so 
that $r(t) = r_{0} \neq 1, \ \forall \ t>0$, which implies 
that $\rho_{D} = \rho_{A}$, $\mu = \chi = \pi (1 - 
\sqrt{1-r_{0}^{2}})/4 \equiv \zeta$, and $\kappa = 
\varphi (1 - \sqrt{1-r_{0}^{2}})/2$. Inserting this into 
Eq. (\ref{eq:fjek}) we obtain  
\begin{eqnarray}
 & & \phi_{g} = 
-\arctan \Big[ r_{0} \Big( 
\frac{\sin \kappa + \sin [\varphi - \kappa] \tan^{2} \zeta}
{\cos \kappa + \cos [\varphi - \kappa] \tan^{2} \zeta} 
\Big) \Big] .
\nonumber \\ 
\label{eq:uhlunitary}
\end{eqnarray}
This case has also been analyzed in the context of 
Mach-Zehnder interferometry in \cite{sjoqvist00}, where it 
was shown that the interference pattern for $r_{0}>0$ 
\cite{remark2} is shifted by the mixed state geometric phase 
\begin{eqnarray} 
\gamma_{g} = -\arctan \Big[ r_{0} \tan \frac{\varphi}{2} \Big] , 
\label{eq:interunitary}
\end{eqnarray} 
where the solid angle enclosed by the Bloch vector is $\varphi$. The
mixed state geometric phases in Eqs. (\ref{eq:uhlunitary}) and
(\ref{eq:interunitary}) match only if $\tan \zeta = 1$ or $\kappa -
\varphi /2 = n \pi$, $n$ integer. The former case corresponds to pure
states and holds for any $\varphi$. As $0< \kappa \leq \varphi /2$ for
$r_{0}>0$, the latter case has solution only for $n=0$ so that $\kappa
= \varphi /2$, which again corresponds to pure states.  Thus, for the
present evolution, $\phi_{g}$ and $\gamma_{g}$ only match for pure
states. In fact, one may show that
\begin{eqnarray}
\left| \frac{\sin \kappa + \sin [\varphi - \kappa] \tan^{2} \zeta}
{\cos \kappa + \cos [\varphi - \kappa] \tan^{2} \zeta} \right| 
& \leq & \left| \tan \frac{\varphi}{2} \right| 
\nonumber \\ 
\Rightarrow \left| \phi_{g} \right| & \leq & \left| \gamma_{g} \right|
\label{eq:relation} 
\end{eqnarray}
with equality in the limit of pure states. This suggests 
that the mixed state geometric phase proposed in \cite{sjoqvist00} 
is more robust to reduction of the length of the Bloch vector 
than that proposed by Uhlmann. 

\item {\it Pure initial state:} This case is characterized by 
$\alpha = \beta = \frac{1}{2}$, which yields  
\begin{eqnarray}
 & & \phi_g = -\arctan  \Big[ \frac{\sin \kappa + 
\sin [\varphi - \kappa] \tan \chi \tan \mu}{\cos \kappa + 
\cos [\varphi - \kappa] \tan \chi \tan \mu} \Big] . 
\nonumber \\ 
\label{eq:pureinitial}
\end{eqnarray}
Thus, from Eq. (\ref{eq:relation}) it follows that 
$|\phi_{g}| \leq \varphi /2$ for pure initial states. 
This reduction of the geometric phase value as a 
function of the decoherence efficiency parameter 
$\Gamma /\omega$ is illustrated in Fig. 2 in the 
case where $\varphi = \pi /2$. We see that the geometric 
phase decreases rapidly with increasing decoherence rate. 
In particular, we note that when the decoherence rate is 
as large as the precession time-scale (i.e. $\Gamma / 
\omega =1$), the geometric phase has practically vanished
\cite{pati96}.  

\begin{figure}[ht!]
\begin{center}
\includegraphics[width=8 cm]{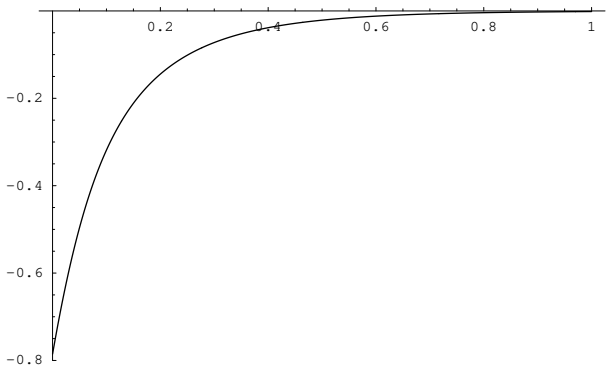}
\end{center}
\caption{Uhlmann's mixed state geometric phase (vertical axis) 
as a function of the decoherence efficiency parameter 
$\Gamma / \omega$ (horizontal axis), $\Gamma$ and $\omega$ 
being the decoherence rate and precession frequency, respectively. 
The precession angle around the $z$ axis is restricted to 
$\pi /2$.}
\end{figure}

\item {\it Small decoherence rate $\Gamma /\omega \ll 1$:} 
First, in the case where $r_{0} = 1$ we may expand the 
right-hand side of Eq. (\ref{eq:integral}) to lowest order 
in the small quantity $\sqrt{1-e^{-\frac{8\Gamma}{3}t}}$ 
yielding  
\begin{equation} 
2\omega \int_{t_{k}}^{t_{k}+\Delta \varphi /\omega} a^2(t) dt = 
\frac{\Delta \varphi}{2} + 
{\cal O} \Big[ \sqrt{\frac{\Gamma}{\omega}} \Big] . 
\label{eq:small1}
\end{equation} 
On the other hand, when $0<r_{0} \neq 1$ and  
$r_{0}^{2} (8\Gamma /3) t_{{\text{max}}} \ll 1-r_{0}^{2}$, 
one may take $\sqrt{1-r_{0}^{2} e^{-\frac{8\Gamma}{3}t}} 
\approx \sqrt{1-r_{0}^{2}} + {\cal O} [\Gamma t]$ so that  
\newpage 
\begin{eqnarray} 
2\omega \int_{t_{k}}^{t_{k}+\Delta \varphi /\omega} a^2(t) dt & = & 
\frac{\Delta \varphi}{2} (1-\sqrt{1-r_{0}^{2}}) 
\nonumber \\ 
 & & + {\cal O} \Big[ \frac{\Gamma}{\omega} \Big] . 
\label{eq:small2}
\end{eqnarray} 
Comparison of Eqs. (\ref{eq:small1}) and (\ref{eq:small2}) 
shows that Uhlmann's mixed state geometric phase in the 
case of weak isotropic decoherence is most fragile for 
pure or nearly pure initial states. 

\end{enumerate}

\subsection{Proposed experiment}
To test the above predictions experimentally requires control over the
dynamics of the ancilla \cite{ericsson02a}. This is very difficult in
the Markovian case, where the ancilla typically has a complicated
structure with many degrees of freedom so as to quickly forget the
information acquired from the qubit. However, one could imagine a
few-qubit implementation adapted to the unitary representation of the
depolarization channel, in case of which Uhlmann's geometric phase
could be realized interferometrically.

The general idea behind such a realization is based upon purification 
lift of the path $t\mapsto \rho (t)$ by adding ancilla systems 
in such a way that partial trace over these ancillas is $\rho (t)$.  
Such a measurement scheme may be constructed as follows:  
\begin{enumerate} 
\item[(i)] Represent the ancilla states $|\widetilde{\mu_{a}} \rangle$, 
$\mu_{a} = 0,\ldots 3$, with a pair of qubits so that 
$|\tilde{0}_{a} \rangle = |0_{a}\rangle \otimes |0_{a}\rangle$, 
$|\tilde{1}_{a} \rangle = |0_{a}\rangle \otimes |1_{a}\rangle$, 
$|\tilde{2}_{a} \rangle = |1_{a}\rangle \otimes |0_{a}\rangle$, and 
$|\tilde{3}_{a} \rangle = |1_{a}\rangle \otimes |1_{a}\rangle$ 
and write the initial state Eq. (\ref{eq:purdepol}) as 
\begin{eqnarray}
|\Psi_{A} \rangle & = & 
\left( \sqrt{\frac{1+r_A}{2}} |0\rangle \otimes |0_{b}\rangle \right. 
\nonumber \\ 
 & & + \left. \sqrt{\frac{1-r_A}{2}} 
|1\rangle \otimes |1_{b}\rangle \right) 
\otimes |0_{a}\rangle \otimes |0_{a}\rangle , 
\end{eqnarray}
where $r_A=r(0)$, and $|\psi \rangle = |0\rangle$ according to 
Fig. 1.  
\item[(ii)] Take $\Psi_{A}$ through the depolarization channel 
around the path of Fig. 1 yielding the final state in Schmidt 
form
\begin{eqnarray}
|\Psi_{D} \rangle & = & \sqrt{\frac{1+r_D}{2}} |0\rangle \otimes 
|D\rangle 
\nonumber \\
 & & + \sqrt{\frac{1-r_D}{2}} |1\rangle \otimes 
|D^{\perp} \rangle  
\end{eqnarray}
with $r_D=r([\varphi + \pi]/\omega)$ and $\langle D|D^{\perp} 
\rangle =0$, where $|D\rangle ,|D^{\perp} \rangle \in {\cal H}_b 
\otimes {\cal H}_a$ are normalized and read explicitly 
\newpage 
\begin{eqnarray}
|D\rangle & = & 
\sqrt{\frac{1+r_A}{1+r_D}} \left( \sqrt{1-p_D} 
|0_{b}\rangle \otimes |0_{a}\rangle \otimes |0_{a}\rangle \right. 
\nonumber \\ 
 & & \left. + \sqrt{\frac{p_D}{3}} |0_{b}\rangle 
\otimes |1_{a}\rangle \otimes |1_{a}\rangle \right) 
\nonumber \\ 
 & & + \sqrt{\frac{1-r_A}{1+r_D}}  \left( \sqrt{\frac{p_D}{3}} 
|1_{b}\rangle \otimes |0_{a}\rangle \otimes |1_{a}\rangle \right. 
\nonumber \\ 
 & & \left. -i \sqrt{\frac{p_D}{3}} |1_{b}\rangle \otimes 
|1_{a}\rangle \otimes |0_{a}\rangle \right) 
\nonumber \\ 
|D^{\perp} \rangle & = & 
\sqrt{\frac{1+r_A}{1-r_D}} \left( \sqrt{\frac{p_D}{3}} 
|0_{b}\rangle \otimes |0_{a}\rangle \otimes |1_{a}\rangle \right. 
\nonumber \\ 
 & & \left. +i \sqrt{\frac{p_D}{3}} |0_{b}\rangle \otimes 
|1_{a}\rangle \otimes |0_{a}\rangle \right) 
\nonumber \\ 
 & & + 
\sqrt{\frac{1-r_A}{1-r_D}} \left( \sqrt{1-p_D} 
|1_{b}\rangle \otimes |0_{a}\rangle \otimes |0_{a}\rangle \right. 
\nonumber \\ 
 & & \left. - \sqrt{\frac{p_D}{3}} |1_{b}\rangle \otimes 
|1_{a}\rangle \otimes |1_{a}\rangle \right) , 
\label{eq:dstates}
\end{eqnarray}
where $p_D = \frac{3}{4} ( 1 - r_D /r_A )$. 
\item[(iii)] Perform a unitary transformation under which 
\cite{remark3}  
\begin{eqnarray}
|D\rangle & \rightarrow & 
|0_{b}\rangle \otimes |0_{a}\rangle \otimes |0_{a}\rangle   
\nonumber \\ 
|D^{\perp} \rangle  & \rightarrow & 
|1_{b}\rangle \otimes |0_{a}\rangle \otimes |0_{a}\rangle  
\end{eqnarray}
so as to obtain the state 
\begin{eqnarray}
|\tilde{\Psi}_{D} \rangle & = & 
\left( \sqrt{\frac{1+r_D}{2}} |0\rangle \otimes |0_{b}\rangle \right. 
\nonumber \\ 
 & & \left. + \sqrt{\frac{1-r_D}{2}} |1\rangle \otimes 
|1_{b}\rangle \right) \otimes |0_{a}\rangle \otimes |0_{a}\rangle .  
\end{eqnarray}

\item[(iv)] Expose $|\tilde{\Psi}_{D} \rangle$ to 
$V_{D} V_{A}^{\dagger} \otimes I_b \otimes I_a$ and let 
the resulting state interfere with $e^{i\delta} |\Psi_{A} \rangle$, 
$\delta$ being a variable $U(1)$ shift. The intensity ${\cal I}$ 
reads 
\begin{eqnarray} 
{\cal I} & \propto & 
\Big| e^{i\delta} |\Psi_{A} \rangle + V_{D} V_{A}^{\dagger} \otimes 
I_b \otimes I_a |\tilde{\Psi}_{D} \rangle \Big|^{2} 
\nonumber \\ 
 & = & 2 + 2 \Re \Big[ \langle \Psi_{A} | V_{D} V_{A}^{\dagger} \otimes 
I_b \otimes I_a |\tilde{\Psi}_{D} \rangle e^{-i\delta} \big] , 
\end{eqnarray} 
where  
\begin{eqnarray}
\langle \Psi_{A} | V_{D} V_{A}^{\dagger} & \otimes & 
I_b \otimes I_a |\tilde{\Psi}_{D} \rangle 
\nonumber \\ 
 & = & \frac{1}{2} \sqrt{(1+r_A)(1+r_D)} \, 
\langle 0|V_{D} V_{A}^{\dagger}|0\rangle 
\nonumber \\ 
 & & + \frac{1}{2} \sqrt{(1-r_A)(1-r_D)} \,  
\langle 1|V_{D} V_{A}^{\dagger}|1\rangle 
\nonumber \\ 
 & = & \mbox{Tr} [\sqrt{\rho_{A}} \sqrt{\rho_{D}} 
V_{D} V_{A}^{\dagger} ] . 
\end{eqnarray} 
Thus, Uhlmann's geometric phase for the path in Fig. 1 is realized as
the Pancharatnam relative phase \cite{pancharatnam56} that shifts the
interference oscillations obtained by applying a variable $U(1)$ shift
to one of the interfering states. Such an experiment could in
principle be implemented using, e.g., ion traps or NMR techniques, by
letting three qubits act as the ancilla systems $a$ and $b$ for a
fourth qubit in the reduced mixed state $\rho$.
\end{enumerate}

\section{Conclusions} 
We have computed Uhlmann's mixed state geometric phase for a qubit
affected by the depolarization channel and have described how this
phase could in principle be realized experimentally. A rapid decrease
of this phase with increasing decoherence rate has been
demonstrated. For weak decoherence we have found that Uhlmann's
geometric phase is most fragile for pure or nearly pure states. In the
unitary case, we have also demonstrated that the mixed state geometric
phase proposed by Uhlmann seems to be more sensitive to reduction of
the length of the Bloch vector than that proposed in
Ref. \cite{sjoqvist00}. In this context, it would be interesting to
compare the results of the present analysis with the extension of the
mixed state geometric phase in \cite{sjoqvist00} to the decoherence
case \cite{ericsson02b}. We hope that this work may lead to further
studies of the Uhlmann phase for various quantum channels as well as
to experiments of mixed state geometric phases in the presence of
decoherence.

\section*{Acknowledgments}
We would like to thank Marie Ericsson and Johan {\AA}berg for 
valuable discussions, and Arun K. Pati for pointing out Ref. 
\cite{pati96}. E.S. acknowledges financial support from the 
Swedish Research Council.


\begin{thebibliography}{99}
\bibitem{remark2} $r_{0}$ has to be nonvanishing as the mixed 
state geometric phase $\gamma_{g}$ in \cite{sjoqvist00} is 
indeterminate for degenerate density operators.
\bibitem{remark3} As is apparent from Eq. (\ref{eq:dstates}), 
this transformation depends upon $r_A$ and $r_D$, which therefore 
have to be known.
\end{thebibliography}
\end{document}